\documentclass[final,narrowdisplay]{elsart}
\journal{Physica A}
\usepackage{graphicx}
\usepackage{amssymb}

\begin{document}
\begin{frontmatter}
\title{Consequences of increased longevity on wealth, fertility, and population growth\thanksref{MNTR}}

\thanks[MNTR]{We acknowledge support by the Ministry of Science of Serbia (project OI141035) and the European Commission (FP6 projects EGEE-II and SEE-GRID-2). RK acknowledges support from the Ministry of Education of Spain (FPU grant AP-2004-7058).}

\author[scl]{A. Bogojevi\'{c}\corauthref{alex}},
\ead{alex@phy.bg.ac.yu} \corauth[alex]{Corresponding author.}
\author[scl]{A. Bala\v{z}}, \and
\author[upf]{R. Karapand\v{z}a}
\address[scl]{Scientific Computing Laboratory, Institute of Physics\\
Pregrevica 118, 11080 Belgrade, Serbia}
\address[upf]{Department of Economics and Business, Universitat Pompeu Fabra\\
08005 Barcelona, Spain}

\begin{abstract}
We present, solve and numerically simulate a simple model that describes the consequences of increased longevity on fertility rates, population growth and the distribution of wealth in developed societies. We look at the consequences of the repeated use of life extension techniques and show that they represent a novel commodity whose introduction will profoundly influence key aspects of economy and society in general. In particular, we uncover two phases within our simplified model, labeled as `mortal' and `immortal'. Within the life extension scenario it is possible to have sustainable economic growth in a population of stable size, as a result of dynamical equilibrium between the two phases.
\end{abstract}

\begin{keyword}
Econophysics \sep Wealth distribution \sep Population growth \sep Longevity
\PACS 89.65.-s \sep 89.65.Gh \sep 87.23.Ge \sep 05.90.+m
\end{keyword}
\end{frontmatter}

\section{Introduction}

Constant lifespan is often taken as one of the assumptions in creating and analyzing econophysical and sociophysical models. Today's bio-medical research is, however, uncovering the reasons why organisms live as long as they do. Through modern genetic engineering, the applications of this research are converging on the point of finding practical ways to extend life substantially (and possibly repeatedly) beyond current life expectancy.

The uncovering of the so called ``secret of life" was one of the crowning achievements of the second half of the past century. The discovery of the structure of DNA by Crick and Watson \cite{dna}, and the later successful translation of the DNA code into the language of proteins, fueled the continuing revolution in molecular biology and bio-technology. This revolution is now making it possible to tackle rationally the complementary question of why we die. For the first time we have the option of looking at death from a fact-based \cite{hmd} scientific perspective. The picture that is emerging is quite unexpected \cite{azbel}. We are getting farther from the concept of ``natural death" as an immutable and inevitable end of life \cite{penna,oeppen,arking}.

Physical models played a crucial role in the discoveries that marked the birth of molecular biology \cite{judson}. Similarly, the well-developed machinery for understanding the behavior of complex systems is today being positioned to help understand the mechanisms of death and the extension of life \cite{rauch,coe,masa,oliveira}. Another avenue of research is to try to understand the social and economic implications of the prolongation of human lifespan. In principle, there are two quite distinct paths that one may choose to take \cite{ball}: economics or econophysics. In this letter we approach these problems from an econophysical framework. Using this language we show that it is possible to model and predict some of the far-reaching social and economic consequences of the successful extension of human lifespan that have, up to now, been disregarded both by economists \cite{blanchard,barro} and physicists \cite{arthur,char,mantegna}.

The fact that we all must die has been one of the central points shaping all human societies. Substantial modification or even the removal of this mortality paradigm will necessarily bring about great change in how societies function. It is important to try to anticipate these changes. Successful modeling of these phenomena is not only of practical, but of heuristic value. Many important discoveries, particularly in physics, have followed from analyzing the consequences of modifying key paradigms.
The introduction of a new and extremely sought-after commodity, allowing for the extension of life, would bring about a great change in economy and society in general.

In this paper we present and solve a simple model dealing with the consequences of just such a novel commodity. We study the implications of possible long-term extensions of life on society and its economy. We model the dynamics of social and economic indicators of a society and investigate how the introduction of life extension will influence fertility rates, population growth and the distribution of wealth. For this purpose we propose and analytically solve a simple model. The presented model, when life extension is absent, is related to earlier investigations \cite{coelho,santos}, the main new feature being the introduction of overall economic and population growth. The presented results include conclusions that the population explosion is not a necessary consequence of introduction of life extension commodity, and that it is even possible to have sustainable economic growth in a population of stable size.

\section{Basic model}

Our model tracks through the generations the number and individual wealth of all of the descendants of a specific individual at generation $t=0$ having wealth $m$. We begin by first treating the simpler case of no life extension. Within our model, the life span of each individual consists of three phases: formative years (parents invest in the individual); adult years (individual inherits some initial money, marries, the pair earns some final amount of money, has children); old age (individual lives off his pension and ultimately dies). As a result of these assumptions, dependency ratios are fixed within our basic model, and are constant throughout the population. We track the adult phase of each individual which starts at $t$ and lasts until $t+1$. We assume that individuals inheriting $m$ money choose spouses having the same amount of money, i.e. that the pair starts off with $2m$ initial capital. While this may appear to be a natural assumption it is not obvious if it holds empirically. For example, Dragulescu and Yakovenko \cite{dragulescu} have studied the related phenomena of the earnings of spouses and have shown the earnings to be essentially uncorrelated. It would be interesting to investigate the correlation of inherited wealth of spouses. Within the model presented here we stay with the above simplifying assumption. We further assume that society is numerous enough so that everyone can find a mate. During their working life the pair increase their wealth by a factor $\gamma$, a fixed constant for the whole society satisfying $\gamma>1$. This money is spent on their children and the pair's pensions:
\begin{equation}
\label{basic}
2\gamma m= kC+km'+2P(m)\ .
\end{equation}
In the above equation $k$ is the number of children, $C$ the investment in each child, $m'$ the inheritance of each child. The pension is assumed to be proportional to initial wealth, i.e. $P(m)=\alpha\, m$. The number of children $k$ is assumed to take on the maximal possible value consistent with the rule $m'\ge m$. This is a crucial assumption strongly affecting the model's predictions. It implies that parents have children only if they can assure them equal or better financial start-up compared to what they had. The number of children thus follows from a simple economic criterion. As a consequence, the model leads to a positive relation between fertility and wealth. For this reason, it is obviously not applicable to poor societies. In those societies the choice of the number of children is more strongly related to survival (procreation) and less to expectations of their future wealth. Our ultimate goal is to analyze the effects of life extension on society. For this reason we focus on developed societies in which economic choices play a dominant role, societies having the necessary financial means to purchase life extension. Although many other things influence fertility (e.g. religion, level of education, system of beliefs), the criterion chosen is that of a simplified model seeking to capture the dominant aspect of the relation between fertility and wealth in developed societies.

We will keep track of $m(t)$ and $n(t)$ (money inherited by the descendants at generation $t$ and the number of those descendants). Note that the total number of people in the society at time $t$ is simply $N(t)\propto n(t)/2^t$.

It is easy to see that for non-trivial dynamics we need to further have $\gamma\ge \alpha+1/2+C/2m$, since smaller values of $\gamma$ lead to $k=0$ for all values of $m$. By introducing the critical value $m_*= C/(2\gamma-2\alpha-1)$, as well as auxiliary quantities $K_1=\left[\frac{2m(\gamma-\alpha)}{C+m}\right]$ and $M_1= \frac{2m(\gamma-\alpha)}{K_1} - C$, we can write the solution of the above dynamics as
\begin{eqnarray*}
k&=&\theta(m-m_*)\, K_1\ ,\\
m'&=&\theta(m-m_*)\, M_1\ .
\end{eqnarray*}
Square brackets denote the integer part of an expression. The step function $\theta(x)$ used here equals unity for $x\ge 0$, and vanishes for $x<0$.
\begin{figure}[!t]
\centering
\includegraphics[width=13cm]{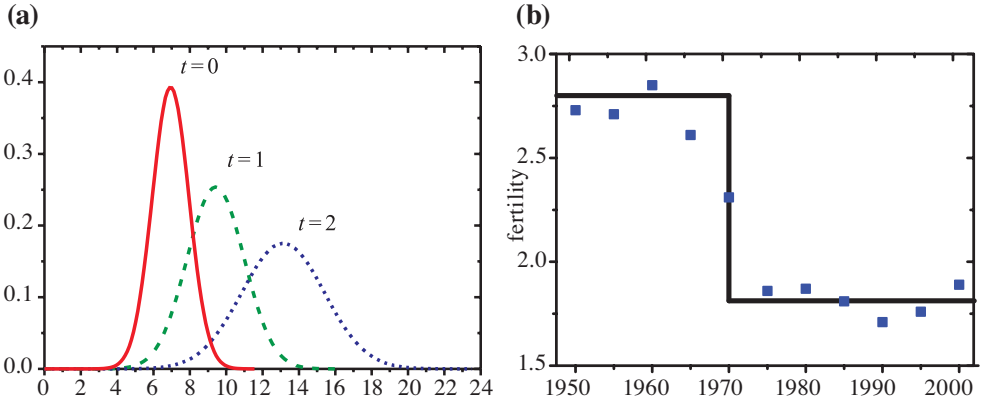}
\caption{(a) Numerical simulations of the time evolution of wealth distributions for the case of no life extension. The plot corresponds to $\gamma=2$, $\alpha=0$, $C=2$, and an initial Gaussian wealth distribution. (b) Comparison of the time dependance of the fertility rate for France \cite{un} with the predictions of the model without life extension when $\gamma-\alpha$ decreases slowly (adiabatically) bellow the critical half-integer value.}
\label{step}
\end{figure}

The above equations make it possible to investigate the dynamical evolution of wealth distributions from given initial conditions. Wealth distributions have been extensively studied in the literature. The field began with the power law distribution of Pareto \cite{pareto}. Recent investigations show that, while Pareto's law gives a good fit at higher incomes, it does not agree well with observed data at middle and low incomes \cite{fuentes,japan,us,uk,chakraborti} which best fit to lognormal or Gibbs distributions. The two regimes follow \cite{silva,fuentes} from the fact that in low and middle income ranges the accumulation of wealth is additive, while in the high-income range wealth grows multiplicatively. Simplified models \cite{trigaux,das,iglesias} have linked Gaussian wealth distributions with egalitarian societies. Our model without life extension agrees well with this phenomenology. The dynamics of the model is such that it preserves Gaussian shaped wealth distributions, as shown in Fig.~\ref{step}a.

In contradistinction to this, Sala-i-Martin has shown \cite{sala,salaweb} that highly segregated societies exhibit bi-modal wealth distributions. In the next section, we will show that the introduction of life extension in our model can lead to societal segregation, which then results in the appearance of just such bi-modal distributions.

Within the framework of our model it is also possible to analyze fertility rates of a given society. The fertility rate $f$ is the average of $k(m)$ over the whole population. If the wealth distribution is such that most of the population have wealth $m\gg C$ then the above solution gives $f\lesssim [2(\gamma-\alpha)]$. Due to the integer part operation, the fertility can depend strongly on small changes in economic growth $\gamma$ or of social expenditures $\alpha$. If $\gamma$ and $\alpha$ change slowly with time, then $\dot\gamma$ and $\dot\alpha$ can be neglected it the equations of motion and we uncover the same relation between fertility, economic growth and social expenditure. As a result, even the smallest decrease of $\gamma - \alpha$ bellow half integer values leads to a decrease of fertility by one unit. This is illustrated in Fig.~\ref{step}b. The data points correspond to measured fertility rates in France. Similar abrupt decreases of fertility have been observed for many other developed countries \cite{un}. Thus, an increase in social expenditures greater than the increase in economic growth results in a step-down in fertility rate. The sharp decline of fertility rates in developed countries has been most often accredited to increased participation of women in the work market. The study of the relation between fertility and wealth within the presented simplified model may offer new insight into this important phenomenology.

It is important to note that the simplifying assumptions made in this paper make the presented toy model (and its generalization to include life extension) analytically solvable. Future models will need to be made more realistic. To do this it will be necessarily to relax some of the assumptions of the current model. In particular they will have to treat the effects of overlapping generations. Unlike in our mean-field model, a more realistic model will have to have agents with different growth factors, different life spans, richer or poorer spouses, etc. We intend to study the effects of the relaxation of these assumptions in a future publication. These more realistic models will have more phenomenological input parameters and will necessitate a purely numerical treatment. We hope that the present analytically solvable model will serve as a useful zeroth-order approximation to these future models.

\section{Model with recursive life extension}

We now generalize the model to include life extension. The extension of life for one individual and one time step (equal to the natural length of the adult period) costs $E$. We assume that any individual having enough money to pay for this life extension will do so, no matter what. Therefore, for $\gamma m\ge E$ we now have:
\begin{equation}
\label{extended}
2\gamma m=2E+kC+km'+2m'\ .
\end{equation}
On the right hand side the first term pays for life extension for the pair, the second and third terms are the investment and inheritance of each child, while the last term represents the ``inheritance" of the original pair with which they begin their new life cycle. Note that we have assumed that both parents inherit the same money as do each of their children. After life extension individuals are assumed to be working able, i.e. there is no pension term in this case. Note that for $\gamma m<E$ we have the same dynamics as before, i.e. as given in Eq.~(\ref{basic}), with $P(m)=\alpha m$. Note that the introduction of life extension decreases dependency ratios - in the extended life periods the population is of good health and is assumed to be in the economically productive phase. Recursive application of life extension drives the dependency ratio to zero.

We introduce $m_1= E/\gamma$ and $m_2= E/(\gamma - 1)$. The life extension phase is for $m\ge m_1$. The solution of the model depends on the relation between critical values $m_*$ and $m_1$. In terms of $K_2=\left[\frac{2m\gamma-2E-2m}{C+m}\right]$ and $M_2=\frac{2m\gamma-2E-K_2C}{K_2+2}$, the solution for $m_*<m_1$ is given by:
\begin{eqnarray*}
k&=&\{\theta(m-m_*)-\theta(m-m_1)\}\, K_1+\theta(m-m_2)\, K_2\ ,\\
m'&=&\{\theta(m-m_*)-\theta(m-m_1)\}\, M_1+\theta(m-m_2)\, M_2\ .
\end{eqnarray*}
Similarly, the solution for $m_*\ge m_1$ equals:
\begin{eqnarray*}
k&=&\theta(m-\textrm{max}\{m_2,m_*\})\, K_2\ ,\\
m'&=&\theta(m-m_1)\, M_2\ .
\end{eqnarray*}
Note that for all values of $m$ the function $\theta(m-m_1)$ measures if life was extended in the current generation.

From the above solutions it follows that, in the life extension phase, $k\ne 0$ is possible only for the case $\gamma> 3/2$, and $m\ge (C+2E)/(2\gamma-3)$. We also see that $m'\ge m$ everywhere except for $m\in[\,m_1,m_2)$. It follows that individuals with $m\ge m_2$ are immortal -- they extend their lives and later they (and their children if they have any) have more money then in the previous life cycle. Note that the potential segregation between mortals and immortals can lead to serious political tensions and instabilities in a society. We next tackle the issue of segregation.

\begin{figure}[!t]
\centering
\includegraphics[width=13cm]{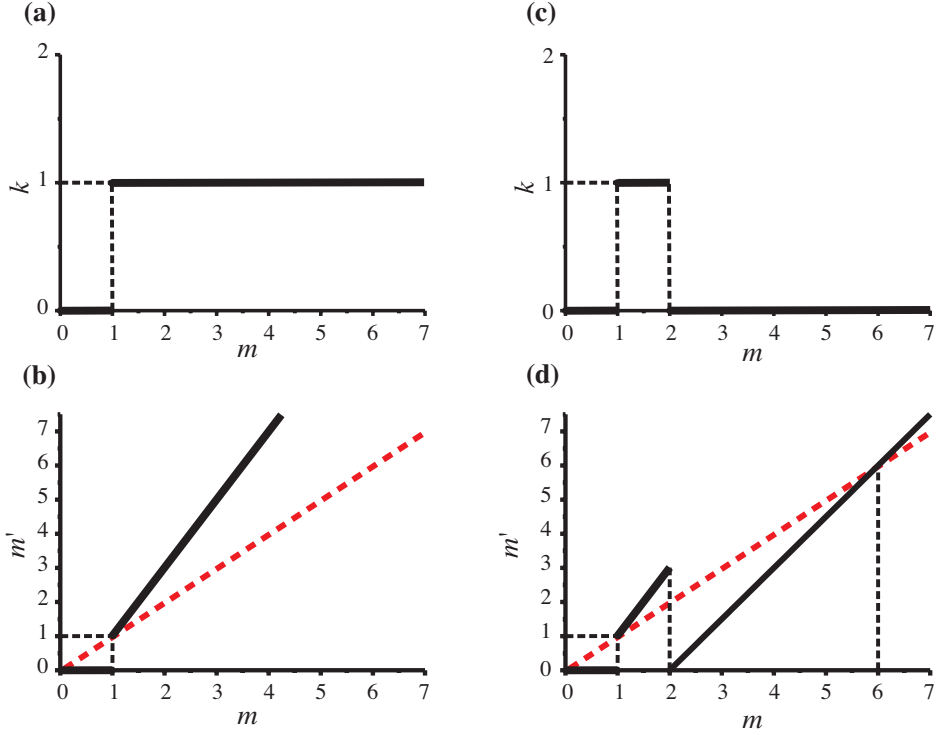}
\caption{Solutions of the model for $\gamma=3/2$, $\alpha=1/2$, $C=1$. Plots (a) and (b) correspond to no life extension, plots (c) and (d) to life extension with $E=3$. The dashed lines in (b) and (d) correspond to $m'=m$. Note that the introduction of life extension has decreased fertility.}
\label{linear}
\end{figure}

Fig.~\ref{linear} illustrates the obtained solutions for the time dependence of the number of children $k$ and wealth $m$ for a society with $\gamma=3/2$, $\alpha=1/2$, $C=1$. Fig.~\ref{linear}a and \ref{linear}b correspond to the case of no life extension. The part of the population with $m<1$ has no children and dies off, those with $m\ge 1$ have one child. That child is financially better off than its parents, i.e. $m'$ is above the dashed $m'=m$ line. After $t$ generations we have $m(t)\sim 2^t$, while the population decreases as $N(t)\sim 2^{-t}$. Fig.~\ref{linear}c and \ref{linear}d correspond to the same society as before but with life extension costing $E=3$. The population now consists of two groups that never mix -- mortals and immortals. Immortals have $m\ge 6$. The number of mortals roughly decreases as $2^{-t}$ while their individual wealth oscillates in the interval $[1,6)$. In fact, numerical simulations show that the majority of this population oscillates in a narrow interval of wealth around $m=2$. For $k=0$ (immortals without children) we have $m'=3m/2-3$, so that their wealth grows asymptotically as $(3/2)^t$. For $k>0$ the condition $m'\ge m$ is not met, i.e. for the case considered immortals can have no children. From Eq.(\ref{extended}) it follows that immortals can procreate only if economic growth is such that $\gamma>3/2$.

Even this single example shows how the introduction of life extension severely affects both population growth and wealth distribution of the whole society, not just of the newly created class of immortals. The key effect of life extension on mortals is that for some of them life span is increased -- some do cross into the life extension phase $m\ge 2$, however, their wealth then decreases (solid curve bellow the dashed $m'=m$ line) and ultimately makes further life extension impossible.

\begin{figure}[!t]
\centering
\includegraphics[width=13cm]{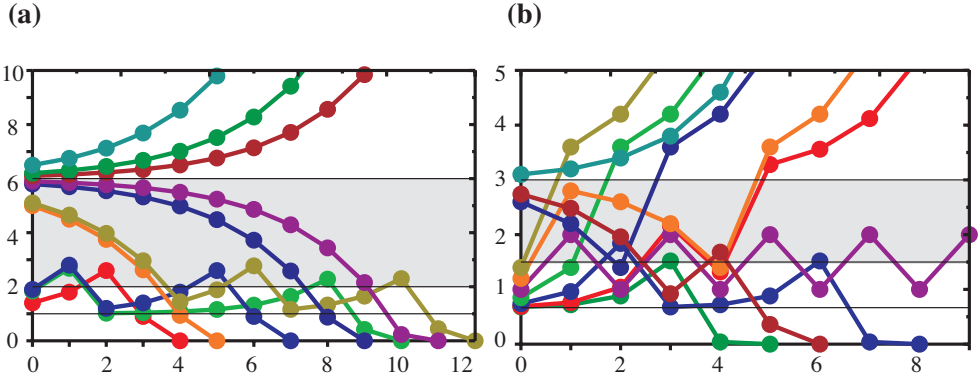}
\caption{Time evolution of individual wealth: (a) An example of a society where tunneling is not possible and mortals and immortals remain segregated ($\gamma=3/2$, $\alpha=1/2$, $C=1$, $E=3$); (b) An example with tunneling ($\gamma=2$, $\alpha=0$, $C=2$, $E=3$). Shaded regions denote corresponding $[m_1,m_2)$ intervals. Horizontal lines beneath them denote the critical value $m=m_*$.}
\label{tunnel}
\end{figure}

Fig.~\ref{tunnel} illustrates the time dependence of wealth for two different sets of parameters. The interval $[\,m_1,m_2)$ represents a barrier through which a mortal must ``tunnel" in a single generation in order to become immortal.
The only way that the descendants of mortals can become immortal is if $m=m_1-0+$ leads to $m'\ge m_2$. To get this we must have:
\begin{eqnarray}
m_* &<& m_1\ ,\\
\frac{2m_1(\gamma-\alpha)}{C+m_2} &\ge& \left[\frac{2m_1(\gamma-\alpha)}{C+m_1}\right]\ .
\end{eqnarray}

Equivalently, these inequalities may be written as:
\begin{equation}
\frac{2\frac{E}{C}(\gamma-\alpha)}{\gamma+\frac{E}{C}\frac{\gamma}{\gamma-1}}\ge
\left[\frac{2\frac{E}{C}(\gamma-\alpha)}{\gamma+\frac{E}{C}}\right]\ge 1\ .
\end{equation}
These inequalities specify a series of isolated islands within the ($\gamma$, $E/C$) plane in which it is possible to tunnel from mortality into immortality. These islands are indexed by integer $n$. For the simplest case $\alpha=0$ they are the areas between curves $\frac{E}{C}=\frac{(n+1)\gamma}{2(\gamma-\alpha)-n-1}$ and $\frac{E}{C}=\frac{n\gamma(\gamma-1)}{2(\gamma-\alpha)(\gamma-1)-n\gamma}$. The critical points $\gamma_n=(1+\sqrt{1+2n(n+1)})/2$ denote the start of the $n$-th island. It follows that mortals and immortals necessarily form segregated populations if $\gamma$ is smaller than the golden mean $\gamma_1=(1+\sqrt{5})/2$.

\begin{figure}[!t]
\centering
\includegraphics[width=13cm]{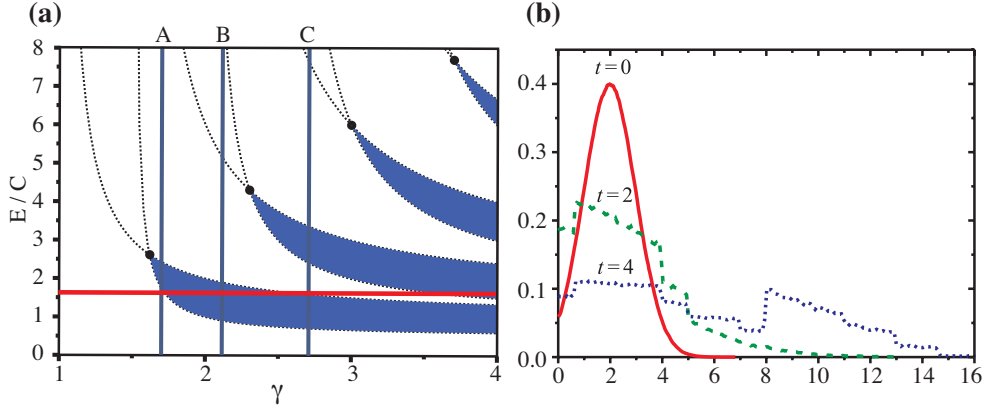}
\caption{(a) Phase diagram showing the isolated islands in the model's parameter space where it is possible to tunnel from mortality into immortality in the case $\alpha=0$. The vertical lines denote $\gamma$ values for: UK, Sweden, Australia (A); US, France, Italy, Canada, India (B); Spain, Greece, Romania (C). For $E/C=1.5$ all the above countries lie on the first island. (b) Time evolution of wealth distributions for the case of life extension with tunneling ($\gamma=2$, $\alpha=0$, $C=2$, $E=3$). In this case the introduction of life extension leads to the emergence of a bi-modal distribution of wealth.}
\label{phase}
\end{figure}

Fig.~\ref{phase}a shows the first few islands in which tunneling is allowed for the case $\alpha=0$ (similar graphs follow for other values of $\alpha$). The vertical lines in Fig.~\ref{phase}a denote $\gamma$ values for: UK, Sweden, Australia (A); US, France, Italy, Canada, India (B); Spain, Greece, Romania (C). These have been calculated using data for adjusted annual growth of these countries from 1960 to 2000 and assuming that one generation in our model corresponds to 30 years \cite{barro}. For $E/C=1.5$ all the above countries lie on the first island in the above phase diagram. The much higher growths of China and Singapore are also consistent with tunneling between mortals and immortals for the same value of $E/C$ since they lie on the second island. Fig.~\ref{phase}b shows that life extension profoundly influences the distribution of wealth. For the society with $\gamma=2$, $\alpha=0$, $C=2$, and $E=3$, tunneling takes an initial Gaussian wealth distribution into a bi-modal one. Wealth distributions of this type are very similar to those of existing highly segregated societies in which life expectancies at birth differ significantly between the rich and the poor \cite{sala}. This is an indication that the presented model, although substantially simplified, captures key aspects of realistic processes.

Similar changes may be found when looking at fertility rates of mortals and immortals, as well as the overall size of the population. In contradistinction to what one might naively expect, the introduction of life extension does not speed up population growth. In fact, for realistic values of $\gamma$ the size of the population generally stabilizes. This gives us interesting examples of societies with sustained economic growth but without a spiraling population explosion. It is not difficult to see that this is a consequence of the dynamical equilibrium between two phases in the model. In fact, this uncovered non-trivial behavior within a simplified model is the essence of how physics can contribute to our understanding of economy and society in general. Namely, effective models in physics (e.g. the Ising model) are of value not because they encode the detailed phenomenology, but because they capture key qualitative relations between dynamical quantities like the one above, providing insight needed for deeper understanding of the underlying phenomenology.

\begin{figure}[!t]
\centering
\includegraphics[width=11cm]{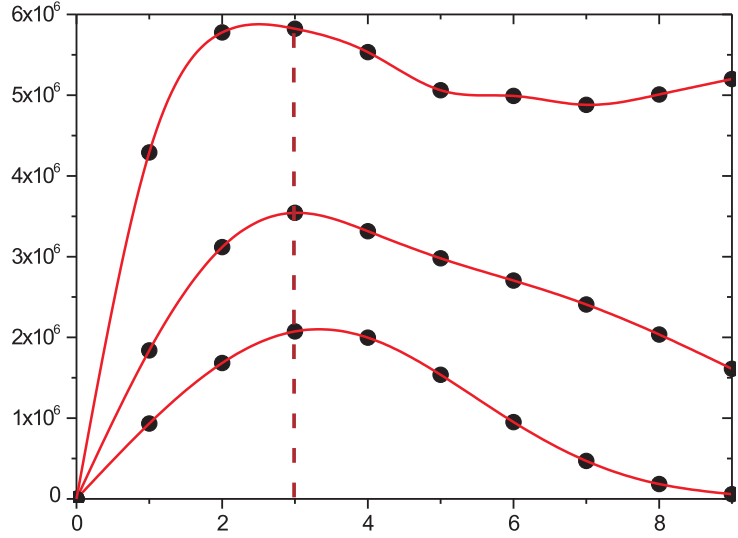}
\caption{Profit of pharmaceutical companies as a function of the unit price $E$. The society shown has $\gamma=2$, $\alpha=0$, $C=2$. The initial wealth distribution was a Gaussian centered at $m=2$ with width $\sigma=1$. The $t=1,2,4$ time slices are shown from bottom to top.  The maximal profit for the producers of life extension is for $E\approx 3$, i.e. for $E/C\approx 1.5$.}
\label{cost}
\end{figure}

So far we have looked at life extension from the consumer's perspective. We now briefly look at the profits of the pharmaceutical companies selling the life extension product. Each individual purchase of life extension increases the profit of the life extension companies by $E$. We assume here that all the R\&D expenses of developing the product have already been covered and that the actual cost of manufacturing the product is negligible. Summing the purchases over the whole society we get the time dependence of the total profit. An example of this is illustrated in Fig.~\ref{cost} for a society with $\gamma=2$, $\alpha=0$, and $C=2$. From the figure we see that the maximal profit determines that $E/C\approx 1.5$. Note that this is also the horizontal line in the phase diagram in Fig.~\ref{phase}a. We see, therefore, that the economic requirement of maximizing profit of pharmaceutical companies is consistent with the political requirement of easing social tension through de-segregating mortals and immortals, i.e. through allowing tunneling into immortality.

\section{Concluding remarks}

We have presented and solved a simplified model that analyzes the consequences of (repeatable) life extension on fertility, population growth and wealth distribution. When life extension is absent the model correctly reproduces observed time dependence of wealth distributions, and abrupt declines in fertility rates. We have analyzed in detail the introduction of life extension to the model and have found it to be a novel commodity which profoundly influences key aspects of society. Of particular interest is the emergence of two distinct phases: societies in which mortals and immortals are segregated, and societies in which economic factors allow descendants of mortals to ``tunnel" into immortality.

The analysis of simplified models such as ours is but a first step in a process that could ultimately help in forming important future policies, e.g. those to do with the pricing of pharmaceutical and medical products and services, wider health-care and insurance policies, etc. As is well known, these issues can have profound effects on the stability of societies and their economic growth, and have for this reason attracted much attention. An important recent example is the decision of the Brazilian government to bypass the copyright on US-produced AIDS drugs \cite{bbc} in order to be able to treat significant part of the AIDS-infected population, and to avoid political instability that may arise from this problem. Models such as ours have the possibility of leading to rationally thought-out policies, allowing society to make critical choices acceptable to its members. However, in order to do this they necessarily need to be followed up by the development and analysis of a series of richer models incorporating more realistic assumptions. We have already commented on some directions in which this process of model building needs to go when we discussed the assumptions within our basic model without life extension. The introduction of life extension will further affect matters such as dependency ratios, or the effects of overlapping generations. The issue of work-leisure tradeoffs can also play an important role in the dynamics of a society with life extension. On the other hand, the very introduction of life extension could greatly influence our attitudes towards work and leisure in prolonged life spans. Also, the decisions based on risk assessment will necessarily undergo a qualitative change when made from longer time perspective offered by repeated life extension. Social and economic strategies will also change accordingly. 

In the presented model the economic growth, parental investment, and price of life extension were all externally determined. More realistic models should include the boot-strap influence of population growth and wealth distribution on these parameters. We plan to pursue this generalization in a future publication. Another interesting extension of the model would be to consider the interaction and co-existence of two parts of society having different parental investments.

\end{document}